\documentclass[manuscript,screen,nonacm]{acmart}

\AtBeginDocument{%
  \providecommand\BibTeX{{%
    \normalfont B\kern-0.5em{\scshape i\kern-0.25em b}\kern-0.8em\TeX}}}

\copyrightyear{2026}
\acmYear{2026}
\setcopyright{none}

\usepackage{graphicx}
\usepackage{array}
\usepackage{colortbl}
\usepackage{xcolor}
\usepackage{enumitem}
\usepackage{algorithm}
\usepackage{algpseudocode}%
\usepackage{makecell}
\setlist{topsep=5pt}

\begin{document}

\title{SenFabric: 3D Printing On-fabric Sensing Mechanisms for Smart Textiles }

\author{Yao Wang}
\affiliation{
  \institution{Graduate School of Design, Harvard University}
  \city{Cambridge}
  \state{Massachusetts}
  \country{United States}
}
\email{yolanda_wang@gsd.harvard.edu}

\author{Mingming Li}
\affiliation{
  \institution{Zhejiang University}
  \city{Hangzhou}
  \country{China}
}

\author{Vera Yu Wu}
\affiliation{
  \institution{Massachusetts Institute of Technology}
  \city{Cambridge}
  \state{Massachusetts}
  \country{United States}
}
\email{verawu@mit.edu}

\author{Marwa AlAlawi}
\affiliation{
  \institution{MIT CSAIL}
  \city{Cambridge}
  \state{Massachusetts}
  \country{United States}
}

\author{Xiaofei He}
\affiliation{
  \institution{Zhejiang University}
  \city{Hangzhou}
  \country{China}
}

\author{Behnaz Farahi}
\affiliation{
  \institution{Critical Matters Group, MIT Media Lab}
  \city{Cambridge}
  \state{Massachusetts}
  \country{United States}
}
\email{behnaz_f@media.mit.edu}

\author{Stefanie Mueller}
\affiliation{
  \institution{MIT CSAIL}
  \city{Cambridge}
  \state{Massachusetts}
  \country{United States}
}

\author{Jiaji Li}
\affiliation{
  \institution{MIT CSAIL}
  \city{Cambridge}
  \state{Massachusetts}
  \country{United States}
}
\email{jiaji@mit.edu}

\renewcommand{\shortauthors}{Wang et al.}

\begin{abstract}

We present SenFabric, a one-step fabrication technique that 3D prints conductive TPU directly onto existing fabric. Rather than compensating for conductive material instability, SenFabric uses geometry-driven resistive structures in which motion changes conductive contact and reconfigures the effective path, producing characteristic signal responses that remain distinguishable despite variation in absolute resistance. We introduce four sensing primitives, slide, press, stretch, and bend, with geometric variables that control activation force, deformation range, and resolution, and a browser-based tool that adapts them to user-supplied outlines and generates routing, pins, and printable models. Four applications demonstrate wearable sensing, human–robot interaction, garment interfaces, and animal wearables, while evaluations show geometry-dominated sensing, cross-motion selectivity, and stable response under repeated use.

\end{abstract}



\begin{teaserfigure}
\begin{center}
  \includegraphics[width=1\textwidth]{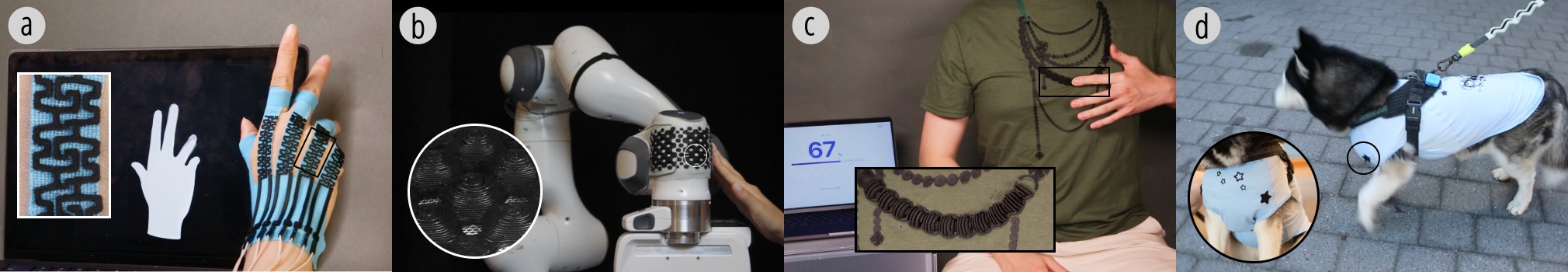}
\end{center}
  \caption{Four 3D-printed objects demonstrating applications of SenFabric, including (a) a hand glove for gesture recognition, (b) a pressure sensor on a robotic arm for directional control, (c) a necklace-shaped sliding interface for volume control, and (d) a dog vest with bend sensors for motion sensing.
}
  \vspace*{10pt}
  \Description{}
  \label{fig1}
  \label{fig:teaser}
\end{teaserfigure}

\maketitle

\section{Introduction}
Smart textiles integrate sensing directly into flexible fabrics, combining physical comfort with functional interactivity.
To overcome the limitations that rigid electronics impose on soft body-conforming materials, HCI researchers have developed a wide range of flexible smart textiles, seamlessly embedding sensing into garments~\cite{ZebraSense,parzer2017smartsleeve}, furniture~\cite{FabriCar,FUrniTURE}, and soft everyday objects~\cite{UIST20_Capacitivo, parzer2018resi}.
As fabrication techniques for smart textiles continue to evolve, manufacturing approaches like knitting~\cite{KnitUI}, embroidery \cite{CHI20_EmbroideredPressure}, and thermal bonding~\cite{electrodermis}, have broadened the design and manufacturing possibilities for smart textiles, each offering a distinct trade-off between fabrication cost, design flexibility, and integration with existing fabrics.

With the advent of accessible, higher-precision 3D printing, a growing number of makers have begun printing directly onto fabric, adding personalized patterns and textures while giving worn or discarded garments new life~\cite{makezine2016tulle,makerworld2025fashion}.
In parallel, HCI researchers have leveraged 3D printing to endow physical structures, integrating sensing functionality directly into the fabrication process itself rather than adding it afterward~\cite{li2026xspine,LattiSense,Flexibles}.
Using additive manufacturing to realize smart textiles is a promising design space to explore: a wide range of existing fabrics can serve directly as print substrates, allowing everyday garments to gain sensing without a custom-built substrate.
Despite this potential, designing 3D-printed structures and patterns that directly encode sensing behavior, without sacrificing the flexibility and form factor that make textiles usable in garments, remains underexplored.

\begin{figure}
  \centering
  \includegraphics[width=0.6\columnwidth]{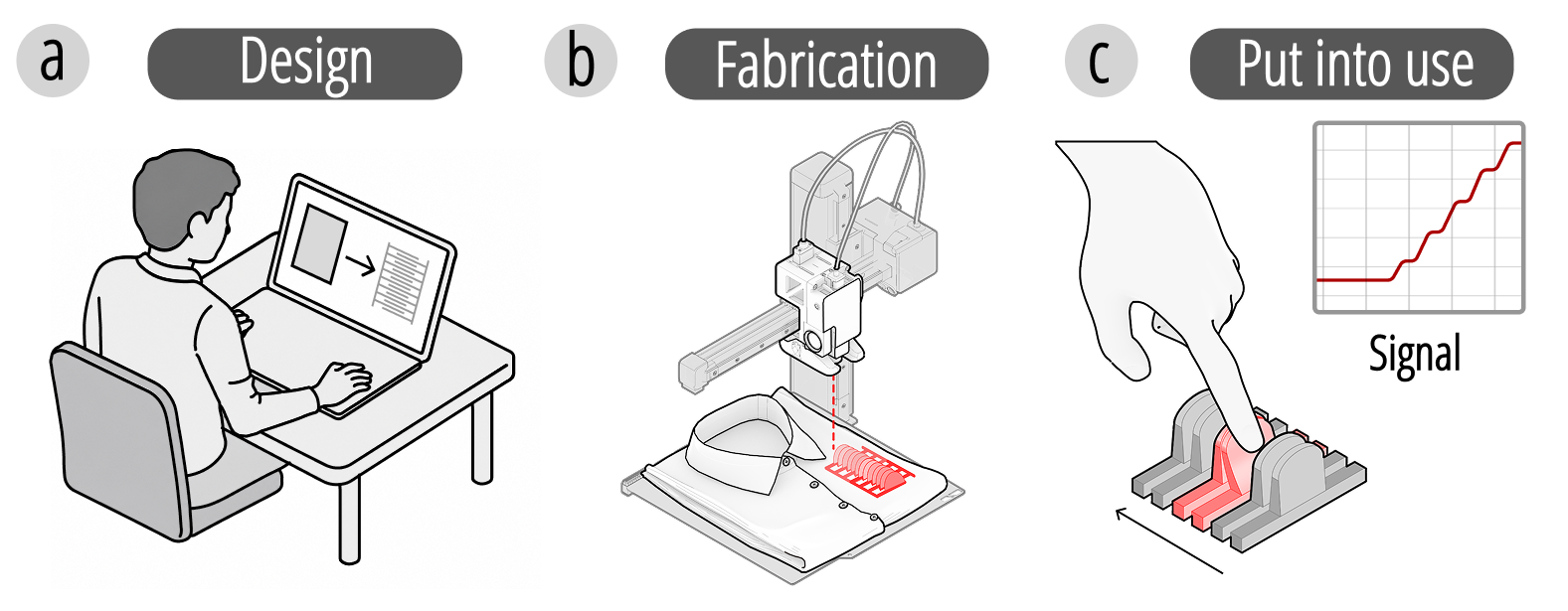}
  \caption{SenFabric's design and fabrication pipeline.
(a) A design tool converts an input pattern into a sensing structure.
(b) The structure is directly 3D printed onto fabric using conductive TPU.
(c) During use, mechanical motion produces corresponding sensing signals.}
  \label{fig:workflow}
\end{figure}

Printing with flexible conductive filament is a promising direction to achieve this goal; however, these materials exhibit time- and history-dependent resistance responses under deformation, limiting the long-term reliability of continuous resistance-based sensing~\cite{DefSense}.
To address this challenge, we present SenFabric (Fig.~\ref{fig:workflow}), a design and fabrication method for directly 3D printing sensing structures onto existing textiles. SenFabric introduces four geometry-driven sensing primitives, slide, press, stretch, and bend, and a design tool for generating corresponding sensing structures from user-defined geometries. We evaluate their sensing characteristics and demonstrate the method through four applications.
By shifting sensing from material deformation to geometry-controlled contact, SenFabric reduces dependence on unstable material responses and enables more reliable sensing on soft, deformable textile substrates.

In summary, our contributions are as follows:
\begin{itemize}
    \item A one-step 3D-printing pipeline to print conductive TPU leads and sensing structures on fabric;
    \item Four sensing primitives (slide, press, stretch, and bend) with geometric variables that control activation force, deformation range, and resolution, and support for combining multiple primitives into a continuous conductive path;
    \item A browser-based design tool that fits any of the four primitives to a user-supplied outline and resolves the routing, the pins, and the printable model automatically;
    \item Four applications demonstrating the versatility of SenFabric across wearable sensing, human-robot interaction, garment interfaces, and animal wearables;
\end{itemize}

\section{Related works}
\subsection{Digital Fabrication for Wearables}

Wearables and smart textiles have integrated sensing through a range of digital fabrication methods, including weaving~\cite{BioWeave,Loopsense}, knitting~\cite{KnitUI,LuxKnit}, embroidery~\cite{CHI20_EmbroideredPressure,TexYZ}, and thermal bonding of conductive sensing components onto existing textiles~\cite{UIST20_Capacitivo,CHI25_IntelliLining}.
These methods provide different fabrication routes for integrating conductive sensing elements into textiles, enabling sensing functionality to be incorporated into garments and other soft textile interfaces.

Digital fabrication has further expanded the design space of wearable systems by directly printing polymers onto fabrics~\cite{RPJ15_DirectPrintTextile}, creating deformable textile-like structures~\cite{UIST19_3DPrintedTextile,UIST20_DefeXtiles,CHI21_ClothTiles,CHI24_FlexTiles}, and investigating fabrication factors such as adhesion between printed polymers and textiles, as well as the softness and stretchability of the resulting structures~\cite{RPJ24_TPUTextile}. These works demonstrate that varying materials and fabrication parameters can produce different physical outcomes in the printed textile structures, demonstrating the potential of 3D printing for customizable textile-based interfaces.

Recent work has also begun integrating sensing structures into digitally fabricated wearables~\cite{UIST22_SenSequins,IrOnTex}. However, the direct use of FDM 3D printing to fabricate conductive sensing structures on existing textiles remains relatively underexplored, particularly in how printed geometry and its interaction with the fabric can be designed to produce diverse sensing behaviors.

\subsection{3D-printed Sensing Mechanism}

Early approaches for integrating sensing capabilities into 3D-printed structures mainly relied on post-processing techniques, such as electroplating~\cite{Intercircuit}, resin injection~\cite{Compressible}, and embedding electronic components~\cite{Fluxable}.
More recently, advances in multi-material 3D printing and conductive filaments have enabled sensing structures to be fabricated directly within the printing process. Prior work has explored conductive structures for capacitive sensing~\cite{Flexibles, MetaSense, MechSense}, resistive sensing~\cite{DefSense, LattiSense}, inductive sensing~\cite{FluxLab}, and contact-based discrete sensing~\cite{li2026xspine}.

From an interaction perspective, researchers have integrated components such as touch sensors, sliders, and touchpads directly into printed geometries~\cite{PrintPut,Capricate}, while other systems support interactions such as pressing, bending, and squeezing through deformable conductive structures~\cite{Flexibles,MetaSense,LattiSense}. Researchers have further explored computational optimization of conductive layouts and circuit geometries to support customizable interaction sensing~\cite{Single-Stroke,network,bae2025computational}.

Prior work has also explored different strategies for improving sensing in deformable 3D-printed structures. DefSense~\cite{DefSense} reported strong time dependence and hysteresis in printable piezoresistive materials, making direct deformation-to-resistance mapping difficult to maintain reliably. LattiSense~\cite{LattiSense} uses lattice structures to amplify deformation-induced resistive changes, while X-Hinges \cite{chang2026xhinges} uses conductive materials with contrasting resistivities to strengthen the contribution of the sensing region and suppress trace-induced noise. In contrast, Xspine~\cite{li2026xspine} converts deformation into discrete contact-state changes through geometric design. Building on this contact-based approach, we explore geometry-driven resistive sensing directly printed on soft textile substrates.

\subsection{Surface Customization of Everyday Objects}

Researchers have explored digital fabrication for customizing the surfaces of everyday objects through geometry, texture, optics, and material properties. For example, Style2Fab~\cite{UIST23_Style2Fab} and TactStyle~\cite{CHI25_TactStyle} support AI-assisted generation of visual and tactile surface features for fabrication. Other work uses optical structures, such as lenticular surfaces~\cite{UIST21_LenticularObjects} and polarized-light patterns~\cite{CHI23_Polagons}, to create viewpoint-dependent or color-changing appearances. These approaches demonstrate how digital fabrication can personalize existing objects by adding new visual and tactile surface properties.
Beyond static appearance, researchers have also explored dynamic and programmable surface customization using color-changing materials, optical effects, and reconfigurable textures~\cite{UIST19_PhotoChromeleon,CHI21_ChromoUpdate,UIST24_PortaChrome}.

While these works primarily focus on visual or tactile customization, SenFabric extends surface customization across three dimensions.
Printed geometries can create regions with different textures and rigidities for tactile customization.
They can also form visual patterns analogous to decorative prints on textiles.
In addition, the printed pattern can encode sensing behavior, with its geometry defining the interaction form factor, such as pressing, sliding, stretching, or bending. In this way, the same printed surface can simultaneously contribute tactile, visual, and sensing functionalities.

\section{SenFabric Method}

SenFabric combines a sensing strategy based on printed conductive geometries with a direct on-fabric FDM fabrication method. This section first introduces the electrical sensing principle underlying SenFabric, and then describes the fabrication strategy used to reliably print conductive sensing structures onto soft textile substrates. Together, these two components provide the foundation for the motion-specific sensing primitives introduced in the following section (Sec. \ref{sec:primitives}).

\subsection{Sensing Principle}\label{sec:principle}

\begin{figure}[H]
  \centering
  \includegraphics[width=\columnwidth]{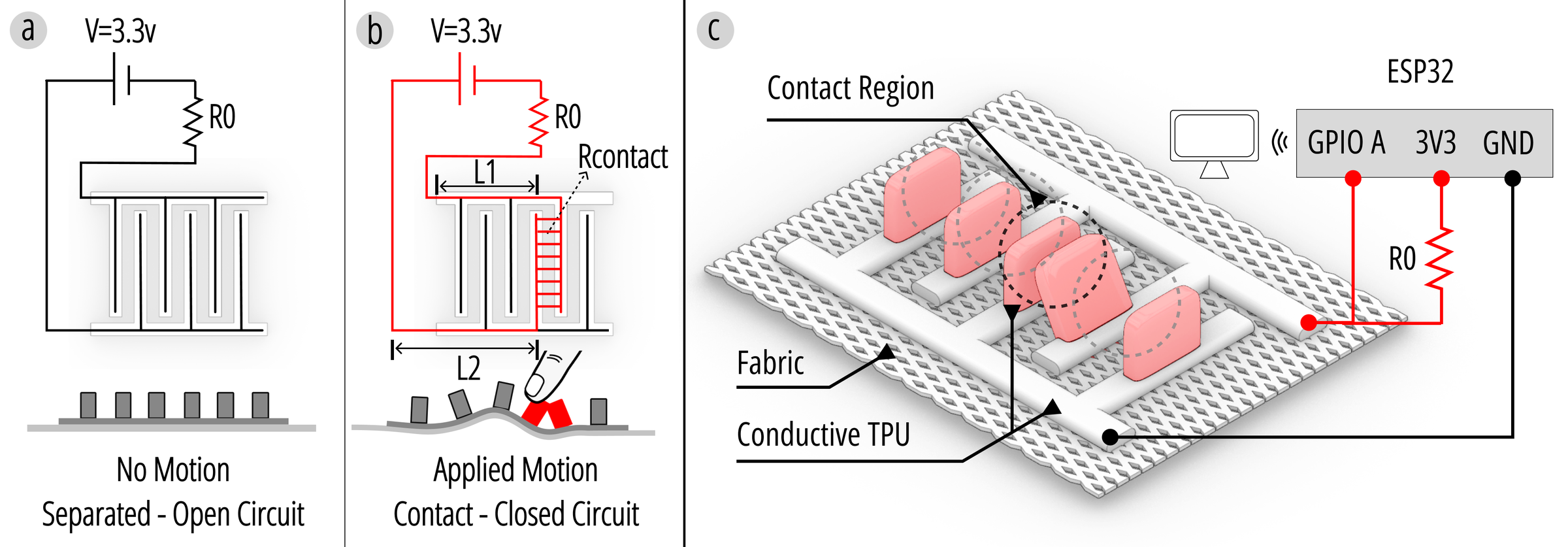}
  \caption{
Sensing principle of SenFabric.
(a) Without applied motion, the conductive electrodes remain separated and the circuit is open.
(b) Applied motion brings the electrodes into contact and closes the circuit.
(c) Printed sensing geometries control where and how many contacts form; different contact configurations produce different equivalent resistances and ADC readings.
}
  \label{fig:sensing_pricipal}
\end{figure}

SenFabric uses geometry-driven resistive sensing based on motion-induced contact between two interdigitated conductive electrode groups. In the resting state (Fig.~\ref{fig:sensing_pricipal}a), the two electrode groups remain physically separated, leaving the sensing path electrically open. When mechanical input is applied (Fig.~\ref{fig:sensing_pricipal}b), the printed structures deform and bring selected regions of the two electrode groups into contact, thereby establishing conductive connections between them.

We describe the resulting electrical state using the contact configuration $C$ between the two electrode groups. The equivalent resistance of the sensing structure can therefore be expressed generally as
\begin{equation}
R_{\mathrm{total}} = \mathcal{R}(C),
\label{eq:sensing_principle}
\end{equation}
where $C$ captures properties of the conductive contact, such as its location, area, and number of contacting regions. Changes in these properties alter the conductive connections and effective current paths within the sensing structure, resulting in different values of $R_{\mathrm{total}}$ (Fig.~\ref{fig:sensing_pricipal}c).

The geometry printed onto the electrodes determines how the contact configuration evolves under mechanical input. Different geometric structures can therefore translate different modes and magnitudes of motion into corresponding changes in contact configuration and resistance. As the applied motion progresses, the contact location, area, or number of contacting regions may change accordingly, producing a characteristic electrical response.

We read $R_{\mathrm{total}}$ using a voltage divider in which a fixed pull-up resistor $R_0$ connects the sensing node to $V_{\mathrm{cc}}$, while the sensing structure connects the node to ground. The output voltage is therefore
\begin{equation}
V_{\mathrm{out}}
=
V_{\mathrm{cc}}
\frac{R_{\mathrm{total}}}
{R_0 + R_{\mathrm{total}}}.
\label{eq:voltage_divider}
\end{equation}
In this configuration, an open or high-resistance sensing state produces an output close to $V_{\mathrm{cc}}$, whereas increased conductive contact reduces $R_{\mathrm{total}}$ and consequently lowers $V_{\mathrm{out}}$.

To maintain useful sensitivity across an expected resistance range $[R_{\min},R_{\max}]$, we select the pull-up resistor as the geometric mean of this range:
\begin{equation}
R_0 = \sqrt{R_{\min}R_{\max}}.
\label{eq:pullup_selection}
\end{equation}
This places $R_0$ at the center of the expected range in logarithmic resistance space and provides a balanced voltage response to relative resistance changes across lower- and higher-resistance sensing states.

Different sensing primitives share this underlying electrical principle while using different printed geometries to translate specific mechanical motions into different contact responses. In this way, SenFabric primarily encodes motion through geometry-controlled changes in electrical contact rather than relying on the intrinsic piezoresistive response of the conductive material alone. The structural designs and motion-specific sensing mechanisms of the four primitives are described in Sec.~\ref{sec:primitives}.

\subsection{Reliable Adhesion in On-Fabric Printing}

Direct 3D printing onto textiles requires both a stable substrate and sufficient mechanical bonding between the deposited material and the fabric. SenFabric achieves reliable on-fabric deposition through four fabrication factors: temporary stiffening of the textile, nozzle--fabric distance, print speed and extrusion, and print-bed temperature.

\begin{figure}
  \centering
  \includegraphics[width=0.6\columnwidth]{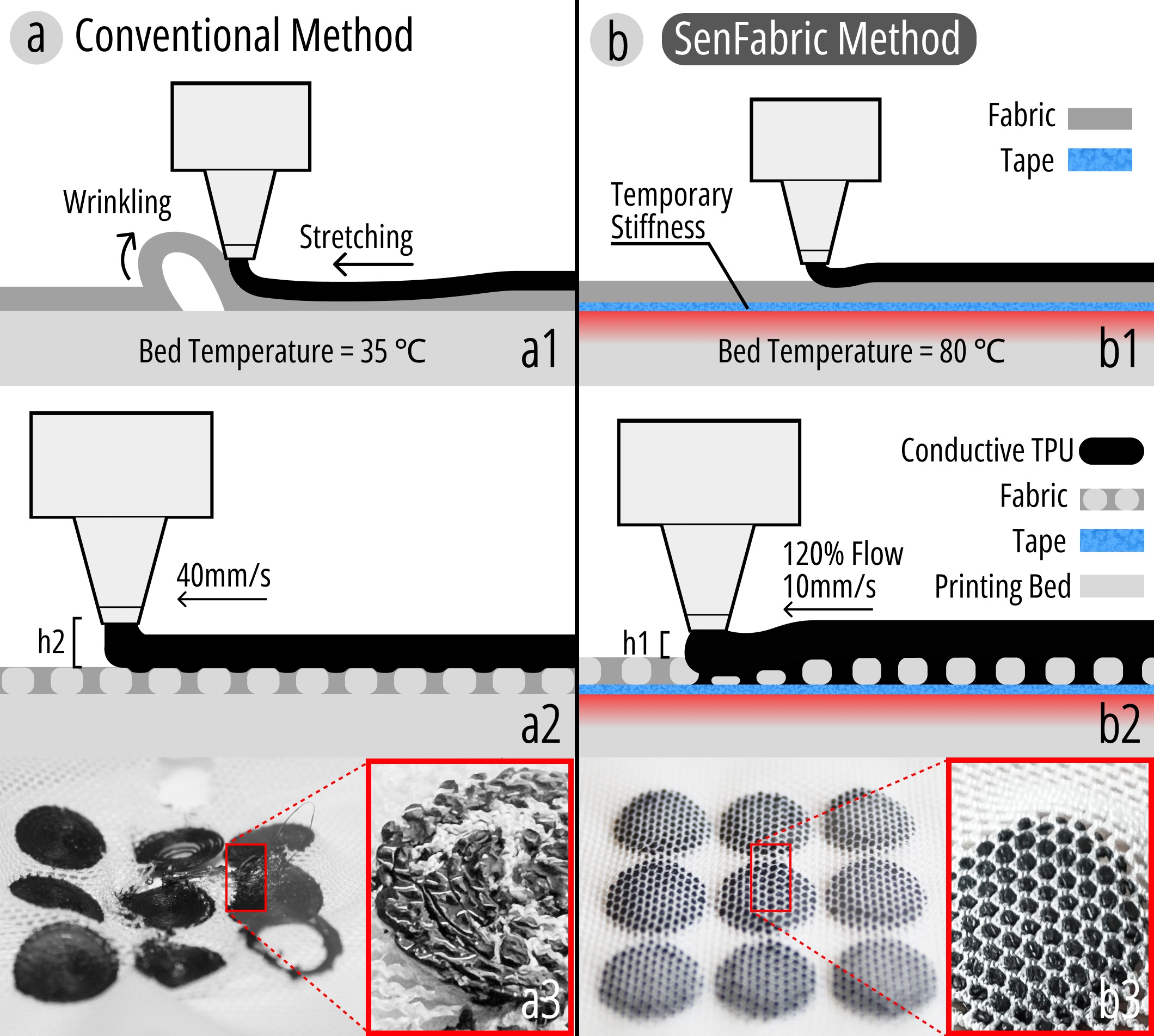}
  \caption{
Effect of temporary tape backing on on-fabric FDM printing.
(a) Without tape backing, the compliant fabric can wrinkle and stretch during printing, leading to uneven deposition.
(b) With temporary tape backing, the fabric is stiffened and stabilized, enabling more uniform conductive TPU deposition and improved print quality.
}
  \label{fig:stiffening}
\end{figure}

\vspace{0.1cm}
\noindent\textbf{Temporary stiffening of the textile:}
Because textiles are compliant, nozzle contact can cause local shifting, wrinkling, or lifting during printing (Fig~\ref{fig:stiffening}a). We attach a thin adhesive backing to the underside of the fabric before mounting it on the print bed (Fig~\ref{fig:stiffening}b), temporarily stiffening the textile and keeping the surface flat and stationary.
The printer is then leveled with the backed textile in place, establishing the nozzle position relative to the actual fabric surface.
We tested this preparation method on mesh, synthetic, and cotton fabrics, and observed reliable printing performance across all three textile types.

\vspace{0.1cm}
\noindent\textbf{Nozzle--fabric distance:}
The vertical distance between the nozzle and textile surface determines how strongly the molten filament is pressed into the weave. A sufficiently small distance promotes penetration between textile fibers and increases mechanical interlocking, whereas a larger distance reduces direct bonding and allows the printed structure greater freedom to move relative to the fabric (Fig~\ref{fig:stiffening}b).
SenFabric uses this difference not only to ensure reliable adhesion, but also to locally tune the mechanical relationship between the printed structure and textile when required by the sensing mechanism, as in the stretch primitive (Sec.~\ref{sec:primitives}).

\vspace{0.1cm}
\noindent\textbf{Print speed and extrusion:}
Print speed $v$ and extrusion rate $\dot{e}$ jointly determine the deposited material volume per unit length, $\lambda=\dot{e}/v$, as well as the time available for the molten filament to penetrate the textile before cooling. Sufficient extrusion and slower deposition increase material contact with the weave and promote mechanical interlocking. In our implementation, we use a print speed of $10$~mm/s for the first layer deposited onto the textile (Fig~\ref{fig:stiffening}).

\vspace{0.1cm}
\noindent\textbf{Print-bed temperature:}
The print-bed temperature further affects adhesion by influencing the cooling and solidification of the deposited filament. Maintaining an appropriate bed temperature slows cooling at the print--textile interface, allowing the molten material more time to conform to and interlock with the textile structure. We therefore maintain a consistent bed temperature throughout fabrication to improve adhesion repeatability (Fig~\ref{fig:stiffening}).

\vspace{0.1cm}
Together, textile stabilization, controlled nozzle distance, material deposition conditions, and bed temperature enable stable and repeatable bonding between printed conductive structures and textile substrates.

\section{Primitives}
\label{sec:primitives}

Building on the sensing principle and fabrication method introduced in the previous section, we further use geometric structure design to enable the printed conductive structures to respond to different modes of mechanical input. These structures follow a common design principle: (1) in the absence of the target input, the two electrode groups remain separated; (2) once the target motion reaches a certain magnitude, the electrodes come into contact and form a conductive path; and (3) as the input magnitude further increases, the contact location, number of contacts, or effective conductive path changes with structural deformation, producing resistance changes that correspond to the progression of the motion.

Fabric surfaces under wear commonly experience four types of local interaction and deformation: tangential sliding along the surface, normal pressure, in-plane tension, and changes in out-of-plane curvature \cite{mcdonald2022interaction,atalay2013textile,gioberto2014garment}. Accordingly, SenFabric introduces four sensing primitives: slide, press, stretch, and bend. In the following, we describe the structural design and sensing mechanism of each primitive. Their characteristic signal fingerprints under target motions, as well as their activation thresholds and selectivity against non-target motions, are quantitatively evaluated in Sec.~\ref{sec:fingerprint}.

\begin{figure}[H]
  \centering
  \includegraphics[width=\columnwidth]{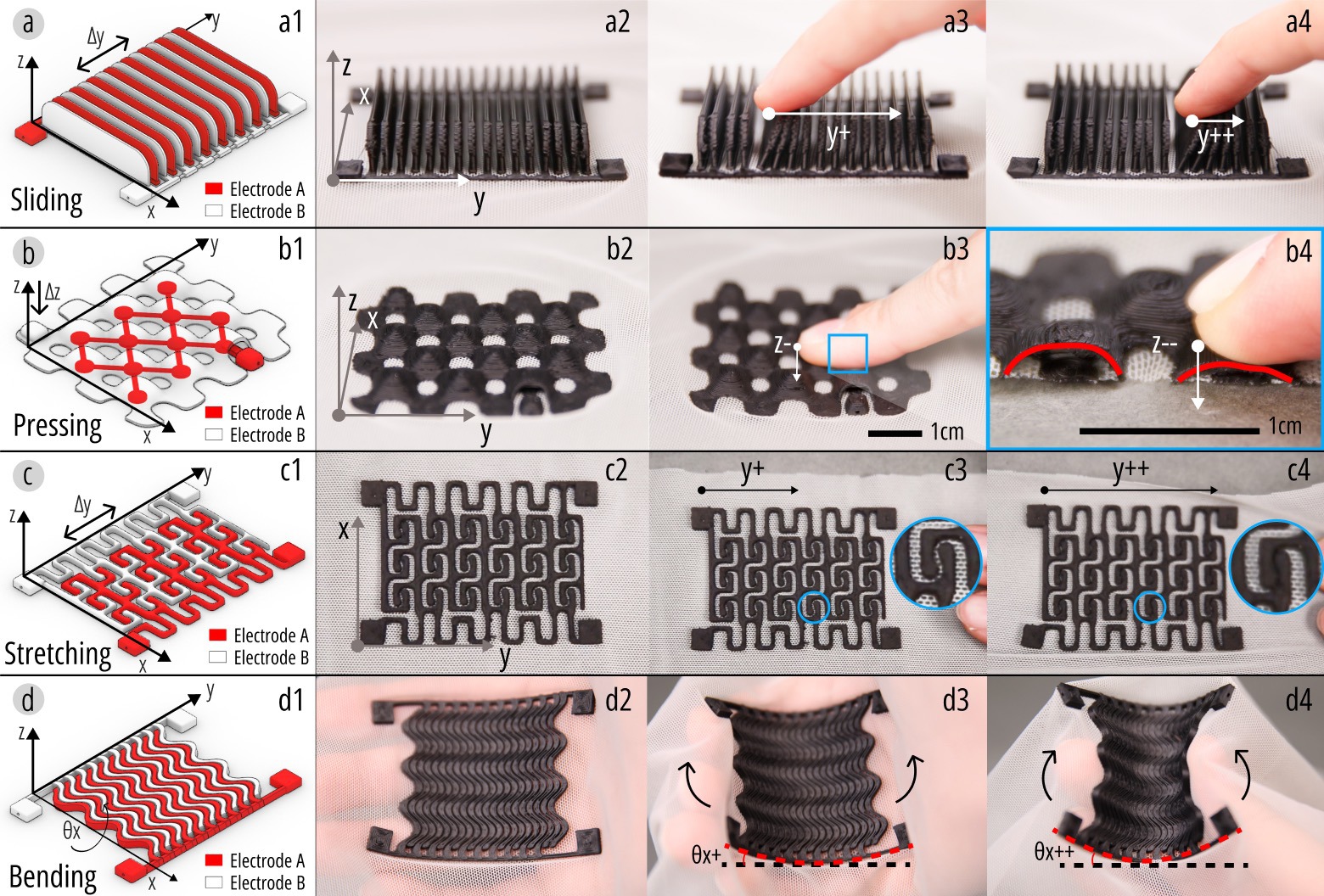}
  \caption{
Sensing primitives of SenFabric: (a) slide, (b) press, (c) stretch, and (d) bend. For each primitive, panel 1 shows the structural design, and panels 2--4 show the fabricated sample under progressively increasing mechanical input from an initial undeformed and separated state to a state of increased electrode contact.
}

  \label{fig:primitives}
\end{figure}

\subsection{Slide (tangential traversal)}

Slide is designed for point-like finger motion along the $y$-axis of the textile surface (Fig.~\ref{fig:primitives}a). It uses relatively tall, thin conductive fins arranged with sufficient spacing to remain separated in the resting state. The increased fin height provides a longer moment arm, allowing tangential finger motion to generate larger bending deformation with a smaller applied force, while also reducing the angular deflection required for neighboring fins to make contact.

As the finger slides along the surface, the fins underneath it are progressively deflected and sequentially brought into contact with neighboring fins from the opposite electrode group. The active contact region therefore propagates along the array with the finger. Different contact locations correspond to different effective conductive paths, indicating the approximate finger position, while the temporal order of successive contacts provides directional information because opposite sliding directions traverse the same contact locations in reverse order.

\subsection{Press (normal pressure)}

Press is designed for sensing normal pressure applied to the textile surface (Fig.~\ref{fig:primitives}b). It uses raised conductive structures positioned above an opposing electrode region, with sufficient separation to keep the conductive paths disconnected in the resting state.
As normal force is applied, the raised structures deform downward and progressively contact the opposing electrode. Increasing pressure enlarges the contact region and reconfigures the effective conductive path, producing a graded resistance response that reflects the progression of compression.

\subsection{Stretch (in-plane tension)}

Stretch is designed for sensing in-plane tension of the textile  (Fig.~\ref{fig:primitives}c). It uses paired hook-shaped conductive features separated by a small initial gap, which keeps the conductive paths disconnected in the resting state while allowing the opposing features to progressively approach each other as the fabric elongates.

The hook regions are printed with reduced adhesion to the textile, enabling limited relative motion between the printed structures and the underlying fabric. As the textile is stretched in-plane, this relative displacement brings the opposing hooks into contact and closes the conductive path. With further stretching, the hooks engage more deeply and the contact region continues to evolve, allowing the resistance response to reflect the progression of textile stretching.

\subsection{Bend (out-of-plane curvature)}

Bend is designed for sensing out-of-plane curvature of the textile (Fig.~\ref{fig:primitives}d). It uses curved conductive features arranged with a small initial separation, allowing neighboring conductive regions to progressively approach each other as the fabric bends while remaining disconnected in the resting state.
As the textile bends out of plane, the curved features conform to the changing curvature and gradually develop a larger contact region. This contact shortens or reconfigures the effective conductive path, and the contact region continues to evolve with increasing curvature, allowing the resistance response to reflect the progression of fabric bending.

\section{Design Space}
\label{sec:designspace}
\begin{figure}
\centering
\includegraphics[width=\columnwidth]{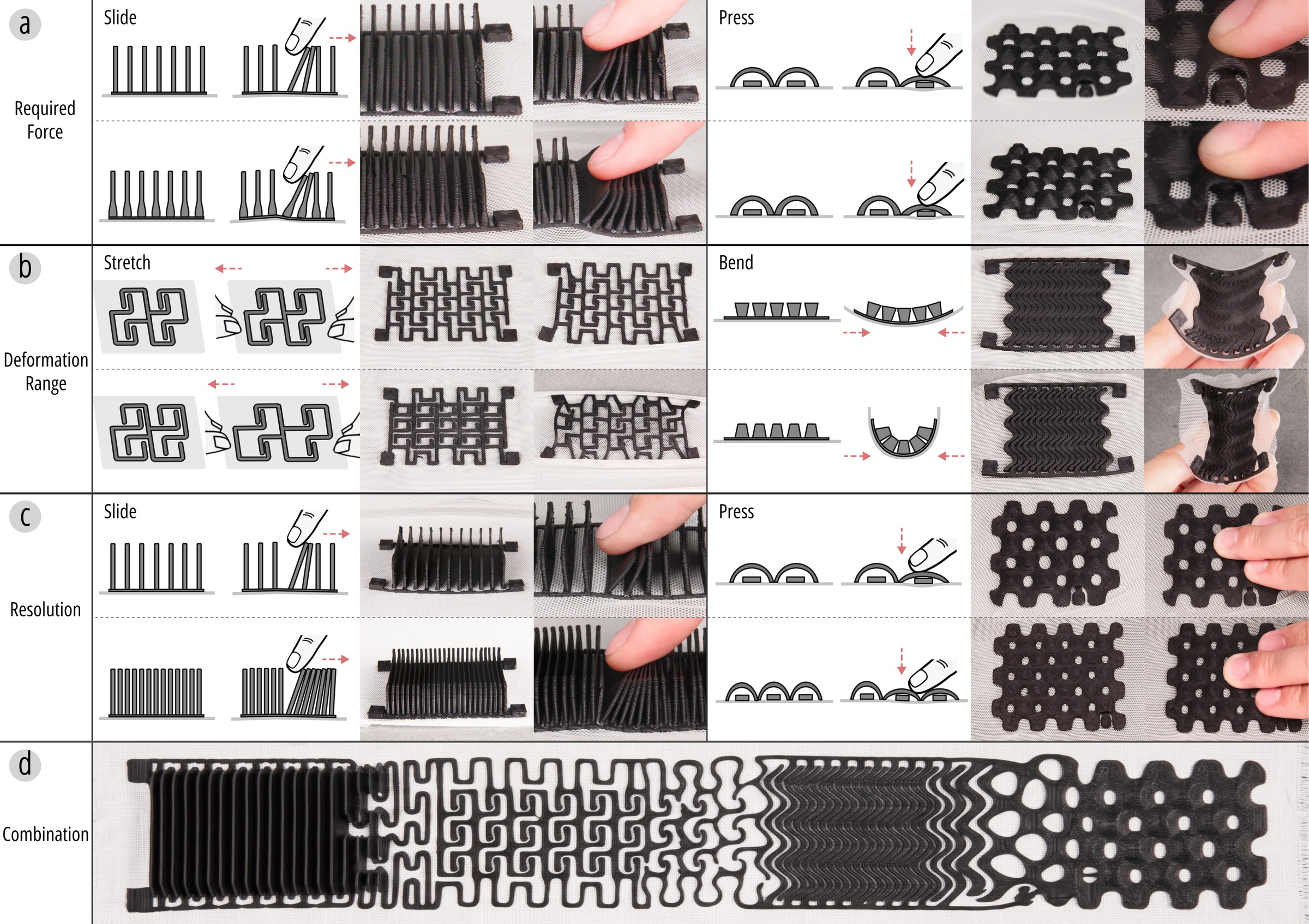}
\caption{Design space of SenFabric. In (a--c) the low configuration is above the dashed line and the high configuration below. Each cell shows schematics at rest and actuated, then the printed sample and its affordance. (a) The required force to deform a primitive can be varied through base thickness. (b) The deformation Range is varied through the gap between adjacent features. (c) The resolution, is varied through feature density. (d) All four primitives can be combined and printed in a single conductive path.}
\label{fig:designspace}
\end{figure}

The four primitives use different printed geometries but share the same underlying principle. Mechanical motion reconfigures conductive contact and effective path length, producing a measurable resistance change (Sec.~\ref{sec:principle}). Three properties of a primitive follow from that change, and we adjust each through one geometric variable. These are the force needed to close a contact, the deformation that closes it, and the number of separate contacts along the way. We demonstrate each variable on two primitives to show that its effect does not depend on a particular sensing structure, and Fig.~\ref{fig:designspace} pairs a low and a high configuration for each case. We then show how primitives can be combined, with the printed pattern carried between them.

\subsection{Required Force}
Activating a primitive requires deflecting a conductive feature until it meets its neighbor. Because the feature is bonded to the textile at its base, the surrounding fabric is drawn into the deflection and resists it. Several aspects of the geometry influence how much force this takes, and we use base thickness as the variable to adjust it. A thicker base is stiffer and is bonded over a larger area, so more force is needed to close the contact. Bond strength per unit area is set during fabrication, not by the printed geometry.

In slide, increasing base thickness from 0.6 mm to 1.2 mm raises the bending stiffness of each conductive pillar and the tangential force needed to deflect it into contact. In press, increasing it from 0.8 mm to 1.5 mm increases the resistance of the raised structure to vertical collapse and the normal force needed to close the contact (Fig.~\ref{fig:designspace}a). In both cases the transition itself is unchanged, since the same states occur in the same number and order.

The two values bracket a practical range, from an input that responds to light touch to one that requires deliberate pressure. The upper end lifts the response above the forces the substrate exerts during ordinary movement, which allows sensing to be placed where the fabric itself bends and stretches.

\subsection{Deformation Range}
When the fabric deforms it carries the conductive features with it, displacing adjacent features relative to one another. A reading appears once that displacement closes the gap between them, and the response saturates once the features are in contact. Releasing the deformation separates them again and returns the signal toward baseline, so the same primitive can be used repeatedly. A textile deforms over a wide range, and we set how much of that deformation is needed before the features meet by changing the size of the gap.

Stretch and bend close that gap in different ways. In stretch, the paired hook-shaped features are carried apart in the plane of the fabric, so their initial gap sets the elongation at which contact begins, and increasing it from 0.5 mm to 6 mm moves the range toward higher strain. In bend, the features rotate rather than translate as the substrate curves, and adjacent features meet along their outer edges. Their center-to-center spacing is the same in both configurations, and only the inclination of the facing surfaces differs. Trapezoids wider at the outer edge leave a small gap there and meet under low curvature, while narrower ones leave a wider gap and require the substrate to bend considerably further (Fig.~\ref{fig:designspace}b).

Because different parts of a garment deform by very different amounts, a knuckle or an elbow far more than the back of a hand, the gap can be set per region rather than once for the whole surface. Each region then keeps its own range, so small movements are not lost where the fabric barely moves and the response does not saturate where it moves most.

\subsection{Resolution}
Resolution is set by the density of conductive features within a given area. Where the response is stepped, the pitch between features fixes the increment in effective conductive path length at each contact and therefore the size of the resistance step between consecutive states. Where the response is graded, density instead sets how finely the contact area grows under load. In both cases a denser array resolves more of the interaction, and the change between adjacent states becomes smaller.

In slide, a finger traveling across the array bridges neighboring pillars in sequence, so the response climbs as a staircase set by the pillar count. Increasing the density from 12 to 24 pillars across a 45 mm sensing region shortens the travel needed to advance one step, so the primitive reports position more finely, provided the spacing stays above what the process can print as separate features. In press, a sparse array of raised structures enlarges the contact between conductive areas in coarse increments as it is compressed, while a denser array does so more gradually and distinguishes more levels of applied force. Our press prototypes use 4 and 6 raised structures per 50 mm (Fig.~\ref{fig:designspace}c).

Density therefore sets how much of an interaction can be read back, and since the states are encoded in the length of a single conductive path, a denser array does so without adding electrodes or read channels. More material over the same area of fabric also changes how the textile feels, so density is raised only where an interaction needs fine input.

\subsection{Combination}
Primitives can be combined in any number and order a surface requires. Adjacent primitives are joined by transitional patterns that keep the printed path continuous across the junction, and a composite is produced in the same single print as an individual primitive. We demonstrate this with all four printed in sequence as one conductive path, read at a single pair of contacts at the end of the strip (Fig.~\ref{fig:designspace}d).

Each primitive in the composite keeps its own geometry, and with it its sensing direction and signal fingerprint. Because each responds to its own mode of deformation and stays below its activation threshold under the others, a deformation acting on more than one of them produces a response built from those fingerprints rather than a single undifferentiated change. Sec.~\ref{sec:selectivity} characterizes this selectivity, including the one case in which bending a strongly deforming substrate can reach the threshold of a low-force press primitive.
\section{Design Tool}
\begin{figure*}[t]
  \centering
  \includegraphics[width=\textwidth]{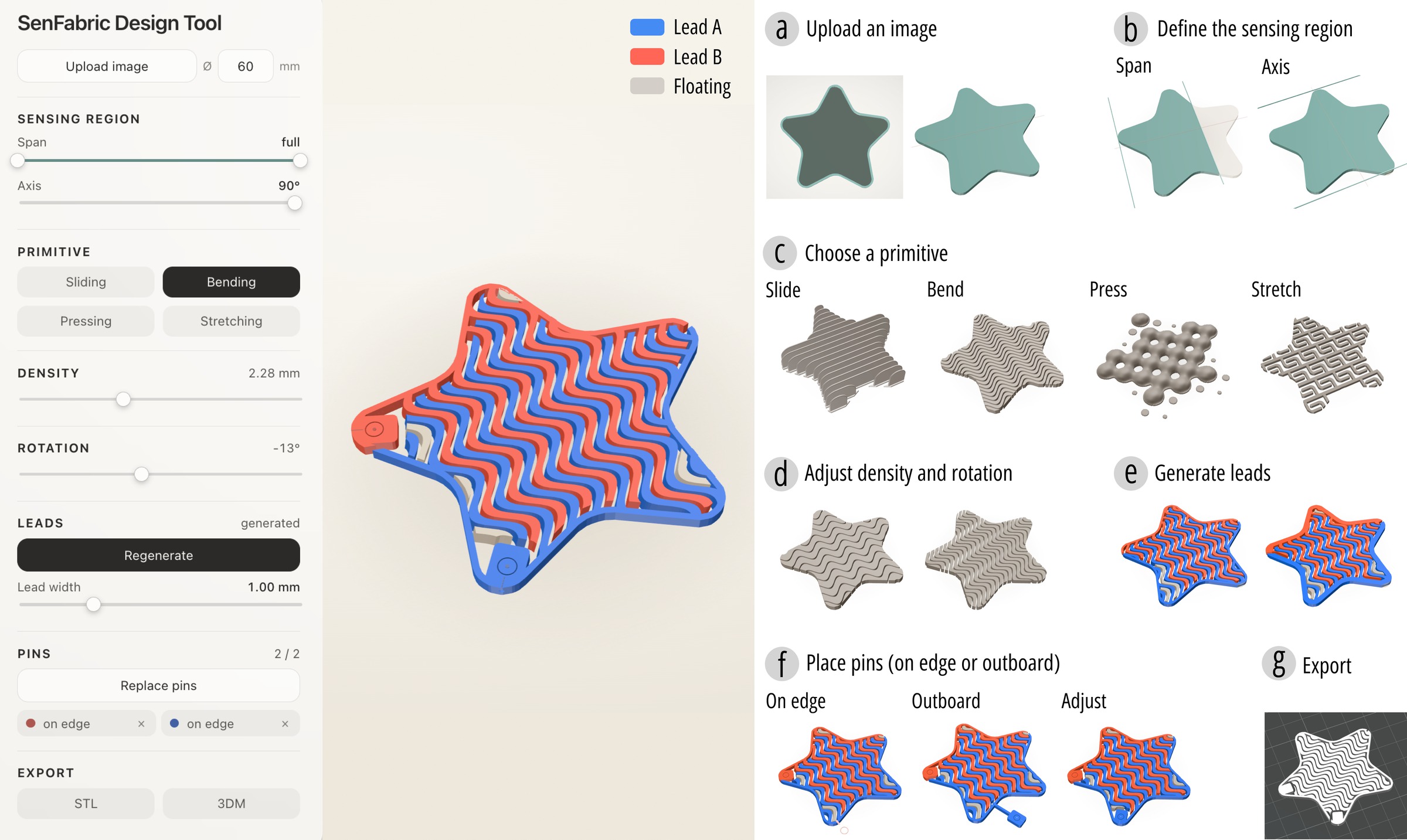}
  \caption{The SenFabric design tool. The interface (left) exposes the controls for
  outline, sensing region, primitive, density and rotation, lead, and pins.
  (a) An outline is traced from an uploaded image and scaled to a user-input diameter.
  (b) (b) The sensing region is defined as a percentage span of the object extent along a chosen axis.
  (c) A primitive is applied to the region. (d) Density and rotation shape the pattern.
  (e) The two conductive traces are generated, either as leads routed from the boundary of the region or, for press, as the layer beneath the shell.
  (f) Two pins are placed, set into the boundary or standing off the part on a stem.
  (g) The model is exported for printing.}
  \Description{Screenshot of the browser-based SenFabric design tool beside seven panels
  showing the stages of generating a printable sensing pattern, from a traced outline to
  an exported model.}
  \label{fig:tool}
\end{figure*}

The SenFabric design tool is a standalone web interface. Operating entirely in the browser, it requires no local installation, backend server, or dedicated CAD license. Pattern generation relies on an exact integer polygon kernel, while 3D solid models are exported via a WebAssembly build of the Rhino geometry library. The user workflow is to upload an image, scale the traced outline, define the sensing region, select a primitive, adjust its density and orientation, generate two conductive leads, place two pin, and export the file (Fig.~\ref{fig:tool}). Each step involves simple slider adjustments or mouse clicks, eliminating the need for manual 3D modeling. With complete rebuilds taking just 65 to 530~ms depending on the primitive, all controls operate live with real-time visual feedback.

Adapting a sensing primitive to an arbitrary shape requires more than geometric fitting. Beyond scaling and clipping, the pattern needs to function reliably as a sensor, which requires two isolated conductive traces, continuous leads reaching every sensing feature, and two dedicated pin attachment points. The tool resolves these steps automatically, with the parameters characterized in Sec.~\ref{sec:designspace} left open to the designer.

\subsection{User Workflow}
\vspace{0.1cm}
\noindent\textbf{Upload an image.} The tool automatically extracts the subject outline from an uploaded image and scales it so that its maximum radius matches a user-specified diameter (ranging from 8 to 400~mm, with a default of 60~mm; Fig.~\ref{fig:tool}a).

\vspace{0.1cm}
\noindent\textbf{Define the sensing region.} The region is given as a span in percent of the object extent together with the axis along which it is measured, and the rest of the outline receives no pattern (Fig.~\ref{fig:tool}b).

\vspace{0.1cm}
\noindent\textbf{Choose a primitive.} Any of the four sensing primitives can be applied to any given outline (Fig.~\ref{fig:tool}c).

\vspace{0.1cm}
\noindent\textbf{Adjust density and rotation.} Density moves the one dimension that is a design variable for that primitive, and rotation turns the pattern within the region without moving the outline (Fig.~\ref{fig:tool}d).

\vspace{0.1cm}
\noindent\textbf{Generate leads.} A click auto-routes the conductive leads, with lead widths customizable from 0.6 to 2.2~mm (Fig.~\ref{fig:tool}e).

\vspace{0.1cm}
\noindent\textbf{Place pins.} The user clicks twice. A click on edge gives a pad set into the boundary and merged with the lead, and a click outside gives a pad standing off the part on a stem (Fig.~\ref{fig:tool}f).

\vspace{0.1cm}
\noindent\textbf{Export.} The generated 3D model is exported for printing (Fig.~\ref{fig:tool}g).

\subsection{Computational Pipeline}
\textbf{Parameter Foundations.} Default geometry settings and parameter ranges for each primitive are derived from reference primitive models. Consequently, generated patterns retain their calibrated physical characteristics regardless of scale or density.

\vspace{0.1cm}
\noindent\textbf{Primitive Generation.} Each primitive is a generative rule over its own measured constants, and the rules are not variants of one another. Slide is two interdigitated combs of arms each carrying a standing wall, bend is two combs that follow the same sine, and stretch is two interlocking hooked combs, and in each case the rule is written so that the clearance between the two traces holds as density changes rather than closing with it. Press is the exception, as its two traces are stacked vertically by placing a domed shell over a flat array of discs and bars. To prevent partial geometry along boundaries, individual dome cells are either fully retained or excluded, creating a halftone-like edge. The same asymmetry reaches the interface, where one pin stands outboard on a stem to meet the base trace while the other seats into the crest of a dome.

\vspace{0.1cm}
\noindent\textbf{Fixed and open parameters.} While design-critical variables like density (Sec.~\ref{sec:designspace}) are exposed to the user, fabrication-constrained parameters, such as the 1.524~mm arm width for sliding patterns, are fixed. Geometric features falling below the 1~mm printing threshold are automatically omitted during generation.

\vspace{0.1cm}
\noindent\textbf{Trace Separation.} For the three in-plane primitives, routing is generated in three phases: \begin{enumerate} \item{Boundary Tracing:} Inward erosion of the sensing boundary creates an initial perimeter band as a closed loop. \item{Parity Splitting:} Features are assigned to one of two traces by band-index parity and trimmed to maintain clearance. \item{Loop Cutting:} The closed loop is severed at its two furthest opposing points, turning the shorted loop into two discrete leads (A and B) that dynamically follow pattern rotation. \end{enumerate}

\vspace{0.1cm}
\noindent\textbf{Verification.} Connected components are recomputed in real time after every edit. Any fragment that cannot reach a lead is marked as floating, providing immediate feedback if an outline geometry prevents complete routing.

\vspace{0.1cm}
\noindent\textbf{Export.} The tool outputs standard 3D models in millimeters as STL files ready for 3D printing and 3DM files containing separate named layers for CAD editing. Contact pads are generated with pre-sized blind bores to receive wiring, requiring no post-processing prior to printing.

\vspace{0.1cm}
For instance, in the dog vest application (Sec.~\ref{sec:applications}), this automated pipeline enables the bend pattern follow a decorative outline while keeping the geometry that gives it its sensing behavior. While the current logic is restricted to two-trace configurations, scaling to multi-channel layouts is a promising direction for future research.

\section{Applications}
\label{sec:applications}

\subsection{Four-Finger Motion Sensing Glove}%
To demonstrate SenFabric for wearable motion sensing, we fabricated a four-finger glove with one stretch primitive printed directly onto the textile along the back of each finger. All four sensing structures are fabricated in a single print without sewing or post-assembly, and are read through four ADC channels on an ESP32. When a finger bends, the textile along its dorsal side stretches locally, activating the corresponding stretch primitive and producing a resistance change that indicates whether the finger is bent or extended.
\begin{figure}[h]
  \centering
  \includegraphics[width=\columnwidth]{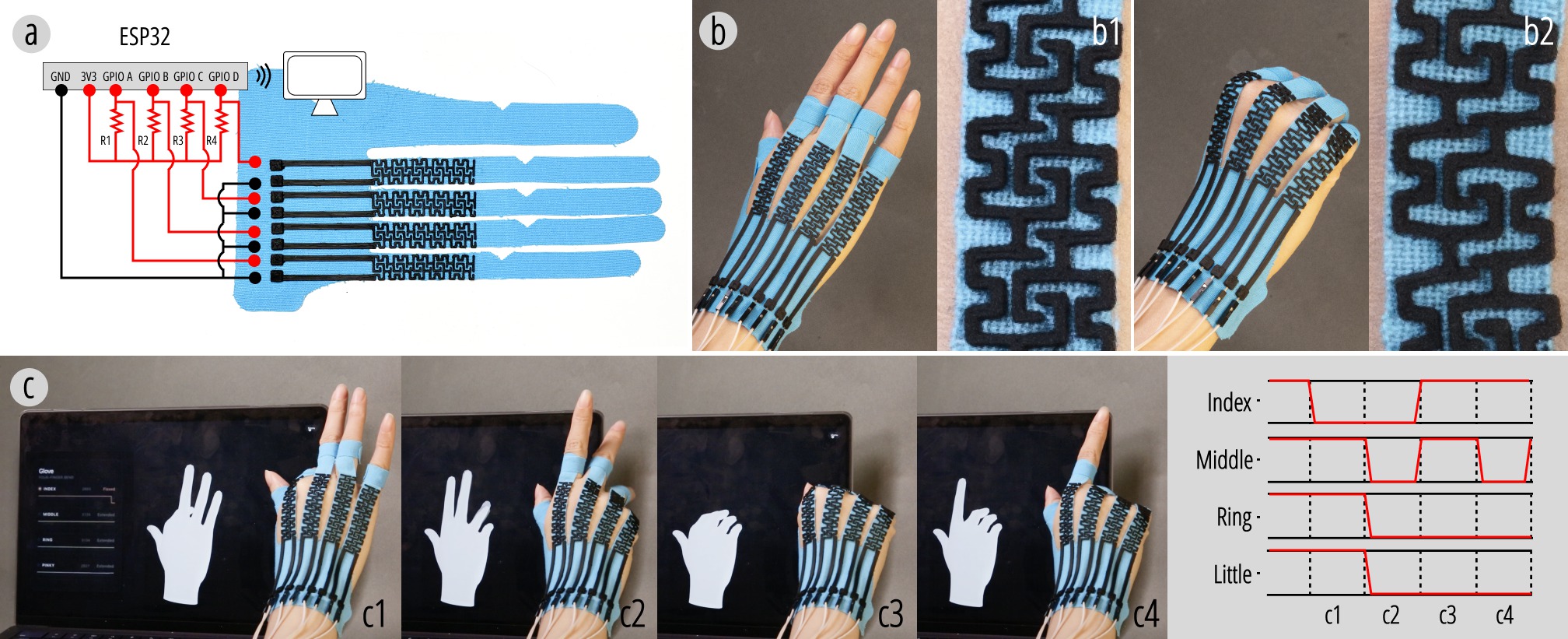}
  \caption{SenFabric enables fabrication of smart wearables like a gesture recognition glove. (a) Each finger on the glove is mapped to a seperate GPIO pin for sensing. (b-c) As the user bends their finger forward, SenFabric electrodes incrementally make contact with neighboring regions enabling smooth motion tracking for each finger.}

  \label{fig:gloves}
\end{figure}

As shown in Fig.~\ref{fig:gloves}, the four channels remain distinguishable during natural wear, including global textile deformation caused by hand flexion and grasping. Different combinations of finger states are mapped in real time to a digital hand model. This example demonstrates that SenFabric can support reliable multi-finger state sensing through sensing structures printed directly onto soft textile substrates.

\subsection{Touch-Sensitive Skin for Direct Arm Control}
In Human-Robot Interaction (HRI), achieving safe, direct close-range collaboration remains a core challenge. Industrial robotic arms offer high precision and payload capacity, but their rigid, insensate shells leave operators with no way to communicate intent through touch, forcing collaboration to fall back on indirect control through screens, joysticks, or teach pendants; limiting a physical interaction that could otherwise be far closer.

\begin{figure}[h]
  \centering
  \includegraphics[width=0.6\columnwidth]{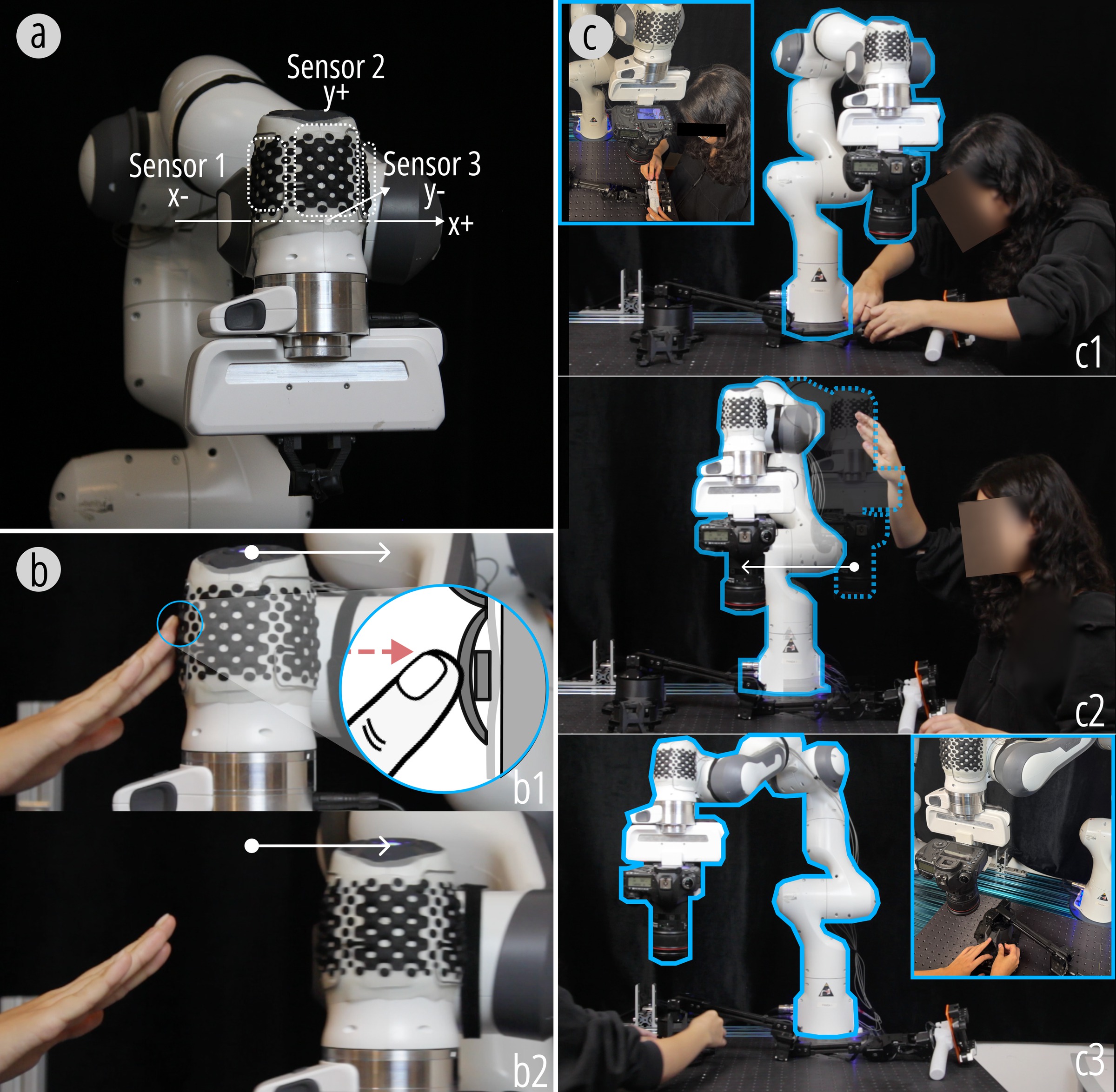}
  \caption{(a) SenFabric press primitives mounted on a robotic arm, enabling human-robot manipulation across the XY axes. (b) A press moves the arm along the axis and direction corresponding to that input.(c) Human and robot as filming partners: a touch on the arm's SenFabric skin guides the camera-equipped arm along the axis, letting the two move in concert to frame the shot.}

  \label{fig:roboticarm}
\end{figure}

SenFabric responds to this limitation at the material and fabrication level: by directly attaching a soft, sensing fabric skin to the arm's surface, it brings touch, an interaction modality widely recognized in HRI but constrained by sensor cost and form-factor adaptation, cheaply onto existing arms, letting operators guide the arm's motion through natural touches and presses, much like resting a hand on a partner's elbow to steer them.

In the camera-assist scenario shown in Fig.~\ref{fig:roboticarm}c, this skin serves as the interface for jointly adjusting the shooting position. A single press on the arm's fabric surface is recognized by direction and force, driving the arm, carrying a 1.8~kg camera, to translate along its axis to the next shooting position. This design couples the arm's motion directly to the operator's touch, letting the shot be repositioned without interrupting the ongoing task or reaching for a separate control interface, echoing HRI's "body-as-interface" paradigm \cite{dahiya2010tactile,rajashekhar2025softissafe}.

This also reflects SenFabric's core design goal for human-robot collaboration: giving an otherwise rigid, insensate arm surface both compliance and sensing, so that safety and expressive interaction become two outcomes of the same fabric layer rather than two subsystems designed separately.

\subsection{Interactive Necklace for Volume Control}

\begin{figure}[h]
  \centering
  \includegraphics[width=0.5\columnwidth]{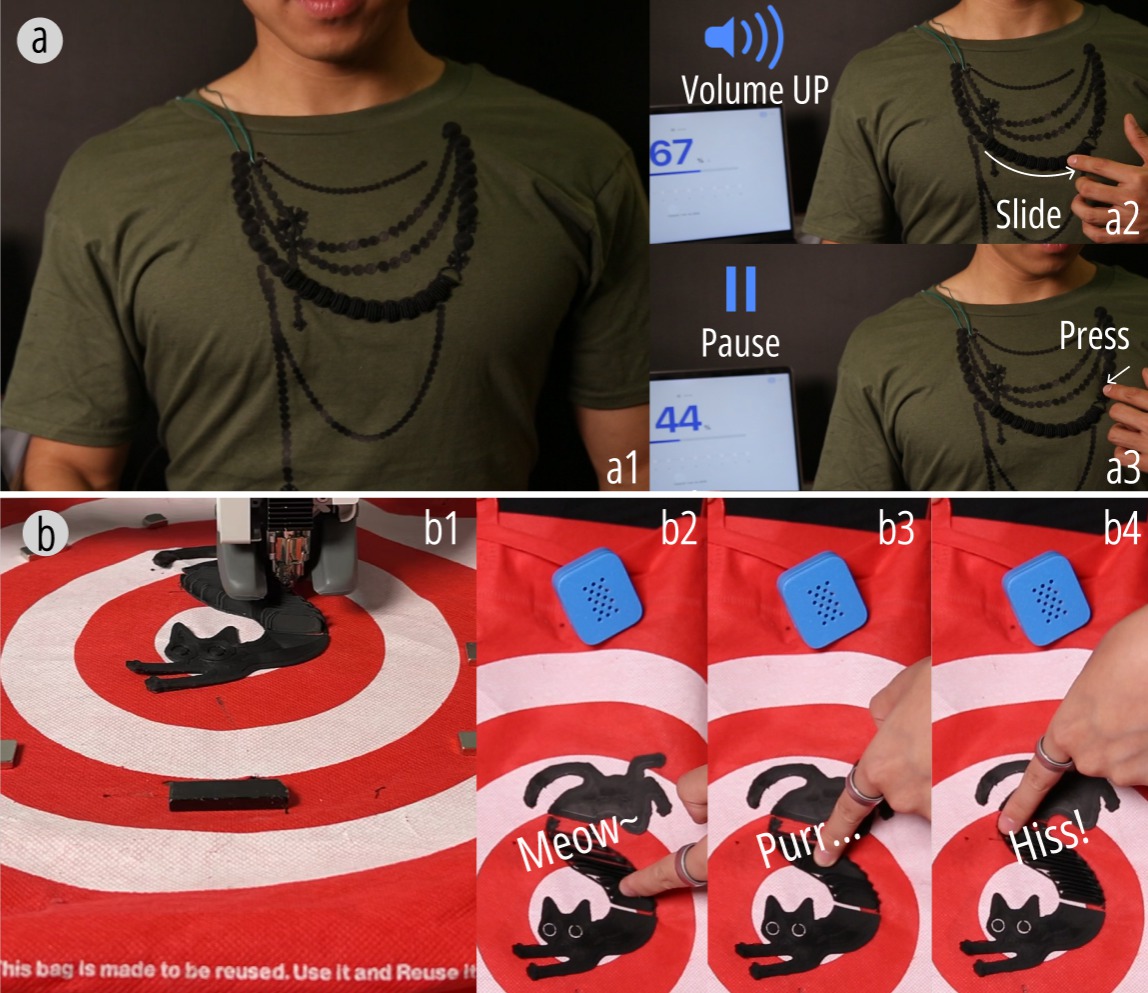}
  \caption{SenFabric turns everyday surfaces into touch interfaces. (a) A necklace-shaped pattern controls audio playback: sliding adjusts volume (a2), pressing pauses it (a3). (b) A printed target bag with a cat motif triggers different sounds, meow, purr, and hiss, depending on where it is touched (b2-b4).}

  \label{fig:tshirt}
\end{figure}

To demonstrate how SenFabric can integrate sensing into decorative elements of garments and everyday objects, combining tactile input with audio output, we printed a necklace-shaped slide-and-press primitive directly onto clothing (Fig.~\ref{fig:tshirt}a). As the user slides a finger along the necklace, successive conductive features come into contact, changing the effective conductive path along the pattern and mapping continuously to volume adjustment; pressing the necklace pauses playback .

We printed a target-patterned press primitive onto a reusable tote bag, with a cat silhouette dividing the sensing region into head, body, and tail (Fig.~\ref{fig:tshirt}b1). Pressing different regions of the cat closes the conductive path of the corresponding structure and produces a location-specific signal, mapped to distinct outputs on a small speaker: touching the head triggers a meow, the body a purr, and the tail a hiss (Fig.~\ref{fig:tshirt}b2-b4).

In both cases, the sensing geometry is embedded within the visual pattern itself, so no separate button, slider, or rigid control interface is required. These applications demonstrate how SenFabric can turn garment and object decoration into functional, spatially selective sensing surfaces while preserving their visual role.

\begin{figure}[h]
  \centering
  \includegraphics[width=0.6\columnwidth]{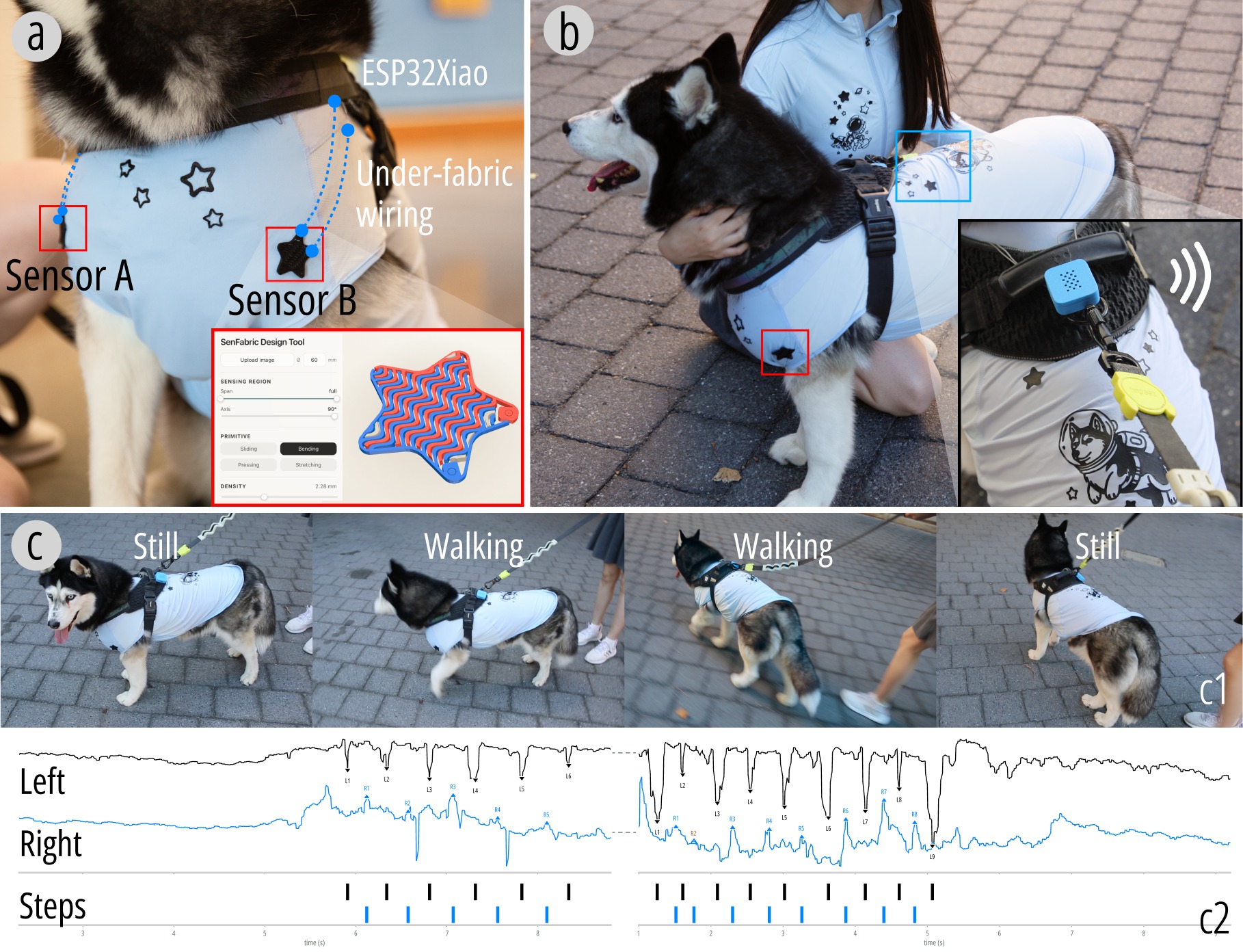}
  \caption{SenFabric for Dog activity monitoring. (a) Two star-shaped sensors integrated into a dog garment and connected to a XIAO ESP32 via under-fabric wiring. sensor geometry is generated from our design tool. (b) Outdoor use with an bluetooth module. (c) Still and walking sequences (c1), corresponding sensor signals, and detected steps (c2). }
  \label{fig:dog}
\end{figure}

\subsection{Activity Sensing Dog Vest}
Wearable sensing is not limited to human clothing. Using our design tool, we fit the bend primitive into a decorative star pattern, producing two bending sensors that take the shape of the design while keeping the primitive's behavior. The sensing structure is absorbed into the decoration, so a vest can be personalized to an owner's taste and still work as a sensor. The two sensors sit over the forelegs of a dog vest, and the traces run underneath the fabric to a Bluetooth module at the collar, leaving no visible wiring and no tethering. As shown in Fig.~\ref{fig:dog}, each stride bends the fabric above a foreleg and produces a transition in that channel. The two channels separate rest from walking, and individual steps can be counted for each leg.

Tracking step counts and activity patterns can help owners understand their dog's daily exercise levels and adjust walking routines. Changes in overall activity or differences between the two forelegs may also draw attention to movement patterns that warrant closer observation.

\section{Technical Evaluation}
\label{sec:Evaluation}

\subsection{Signal Fingerprints}\label{sec:fingerprint}

In this section, we characterize each of SenFabric's primitives (Section 4) in response to their corresponding forces: sliding, compression, stretching, and bending, as a function of change in sampled ADC signal.

To do so, we fabricate our sensing primitives by 3D printing conductive TPU (Conductive Filaflex, Recreus) directly onto a nylon–spandex sheer mesh using a Bambu A1 printer. Each primitive measures 61 × 43 × 10 mm. To characterize the effect of each force, we actuate the primitive on a dedicated mechanical rig, with the sample connected to an ESP32 microcontroller, at two contact points, that continuously records resistance throughout actuation. All metallic fixtures in contact with the sample are electrically isolated using paper shims.

Because each primitive engages a distinct physical interaction,  the test setups and actuation methodologies differ accordingly for each. As a result, we do not evaluate primitives by the magnitude of their response alone. Instead, since the printed geometry reshapes the conductive path on contact, each primitive is evaluated by the morphology of its ADC signal: where the response begins and where it saturates, whether the loading and unloading strokes retrace the same path, and whether the response is graded or resolves into discrete levels.

Next, we summarize our tests of the four primitives under lateral sliding, normal compression, uniaxial elongation, and bending. We also derived an ordinal descriptor for each primitive, used for discrimination in Section~\ref{sec:selectivity}.

\begin{figure}[h]
  \centering
  \includegraphics[width=\columnwidth]{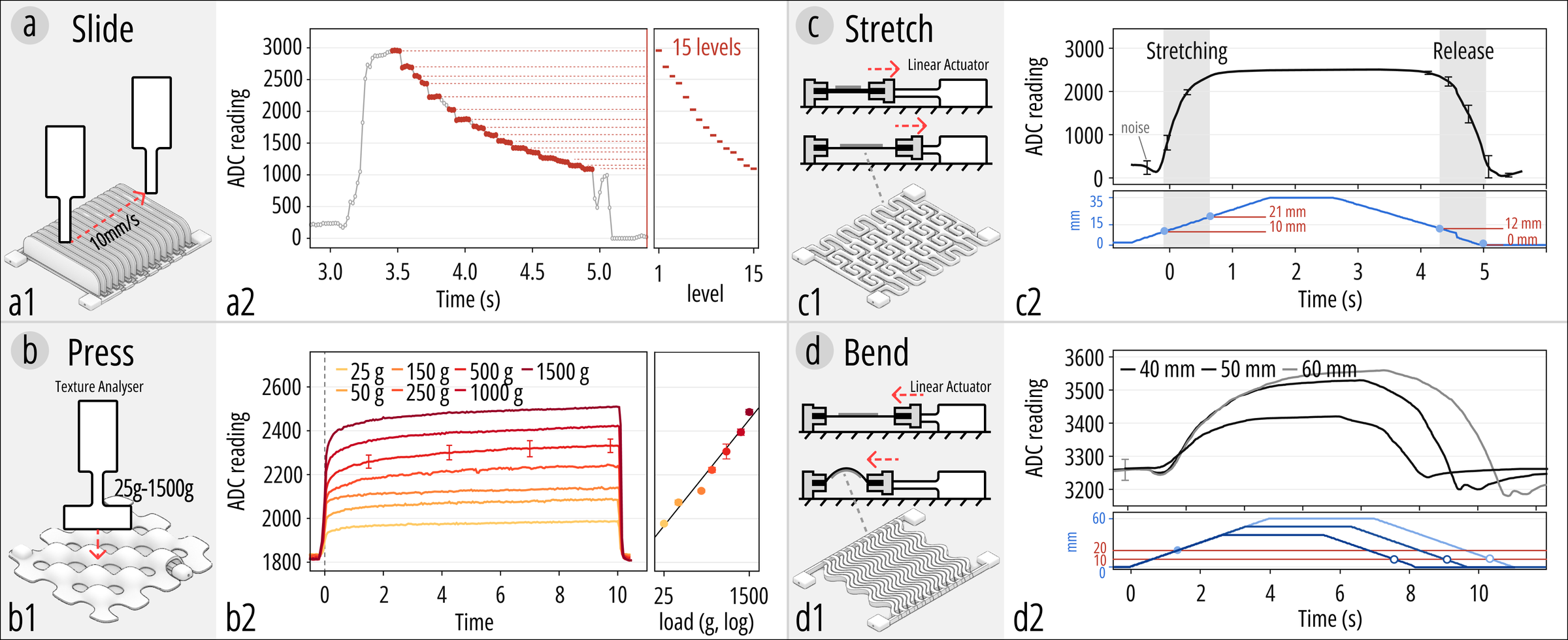}
  \caption{Signal fingerprints of the four primitives. Sampling rates and repeat counts differ by primitive: (a) Slide, 55 Hz, ten passes (five forward, five backward) on one sample; one representative return pass (trial 19) is shown with every raw sample, red marks runs of samples whose successive change stays under 1.2× the baseline noise, and the side plot lists the fitted levels in order. (b) Press, 200 Hz, three presses at each of seven loads; the side plot shows the plateau against load on a log axis with a least-squares fit. (c) Stretch, 1 kHz, ten load–unload cycles. (d) Bend, 200 Hz, five cycles each at 50 and 60 mm and ten at 40 mm; the 40 mm run is a different printed sample. Error bars are ±1 SD at fixed intervals (every 1 s in c, 2 s in d, 2.5 s in b). Lower strips in c and d show actuator position. Absolute ADC levels are not comparable between panels.}

  \label{fig:fingerprint}
\end{figure}

\vspace{0.1cm} \noindent\textbf{Slide} SenFabric slide primitives comprise an alternating array of conductive pillars that engage and disengage in response to an applied lateral force. To characterize the effect of sliding on the resulting signal, we mount a rigid, stylus-shaped end effector on a plotter's pen carriage and traverse it along the primitive at 10 mm/s, with the tip held at a fixed height of 8 mm above the substrate. As shown in Fig.~\ref{fig:fingerprint}a, engagement begins with an abrupt transition of $\Delta$ADC = 620 (SD 111) counts, after which the signal climbs monotonically to a peak of $\Delta$ADC = 2856 (SD 71) before disengaging in a second abrupt transition. The engagement interval is 1896\,ms (SD 26\,ms). The two directions are separated by the position of the peak within this interval, which occurs at 0.90 (SD 0.02) of the interval in the forward direction and at 0.16 (SD 0.01) in the backward direction, with no overlap between the two groups. This is expected because the stylus engages the asymmetric scale profile from opposite ends, so the conductive path lengthens in one direction and shortens in the other. Notably, the baseline ADC reading varies from 56 to 288 counts, yet the normalized peak position remains separable.

\vspace{0.1cm} \noindent\textbf{Press}
Our apparatus comprises a texture analyzer (TA.XTplusC Texture Analyser, Stable Micro Systems) with a load programmed to apply a 10-second compressive force to the compression primitive as seen in Fig.~\ref{fig:fingerprint}b. We ran compression tests spanning 25g--1500g, which corresponds to light human touch (screen-tap) through firm press \cite{displaymodule2015forcetouch,forcerange2016}. For each applied force, we averaged the signal across its three trials and computed the percent change in signal (ADC reading). As shown in Fig.~\ref{fig:fingerprint}b, the relationship between applied force and the change in resistance is logarithmic, following
$\%\Delta_{\text{ADC}} = 15.6906 \log_{10}(F) - 15.2661$
($R^2 = 0.9727$), with 25\,g corresponding to a 7.7\% change and 1500\,g to 36.7\%. Unlike the other three primitives, compression deforms the conductive path continuously rather than reshaping it at discrete contact events, which is why the response is graded rather than stepped.  We therefore describe press by the slope of its log-linear ADC–load relation and by the upward drift of each level during a sustained hold.

\vspace{0.1cm} \noindent\textbf{Stretch} Our setup consists of a stretch primitive (Fig.~\ref{fig:fingerprint}c) clamped at both ends and elongated by a linear actuator (L12-100-100-6-R, Actuonix) over a 60\,mm gauge length. We stretched the sample at 10\,mm/s. As shown in the figure, elongation produces a single monotonic transition rather than a sequence of discrete events. For our primitive samples, the signal remains at the baseline (starting) ADC value through the first 10\,mm (17\% strain), rises steeply between 10 and 21\,mm with a peak sensitivity of 13.6\,\%FS/mm at 10\,mm, and then flattens. This is expected because elongation progressively enlarges the contact area at each interface along the gauge length. The loading and unloading strokes do not retrace the same path: the return stroke stays near the plateau until the sample is almost fully relaxed, reaching 93.2\% of full scale at 8\,mm. Both stretch and bend involve deformation across a large area of fabric, so the enlarged contact areas persist as the sample relaxes. Beyond 23\,mm elongation (38\% strain) the incremental sensitivity falls below the plateau noise ($\sigma$ 17.8 counts) and further elongation is indistinguishable from saturation. The mesh substrate does not reach this strain in normal wear, so the primitive operates within its working range throughout use. Over that range the transition spans $\Delta_{\text{ADC}} = 2202$ counts ($\sigma$ 169).

\vspace{0.1cm} \noindent\textbf{Bend} Finally, to characterize the effect of bending, we built an apparatus that secures the bending primitive to a thin living hinge capable of bend radii smaller than 3\,mm (Fig.~\ref{fig:bulk}). To ensure deformation is controlled during bending, we fixed one end of the primitive to a stationary platform and the other end to a linear actuator (L12-100-100-6-R, Actuonix) that moves along a fixed horizontal rail toward the platform, ensuring purely axial displacement. The total length between the fixed base and the linear actuator's end effector is 110\,mm. To bend the primitive, we extended the linear actuator to distances of 40\,mm (bend radius 105), 50\,mm (bend radius 90), and 60\,mm (bend radius 75), then returned it to its un-extended state, recording ADC readings throughout.
As shown in Fig.~\ref{fig:fingerprint}d, bending produces a single graded transition that stays at rest through the first 20\,mm of actuator travel, rises as the primitive continues to bend, and flattens once the hinge reaches its set displacement. Deeper bending raises the amplitude without moving the transition: the response spans $\Delta_{\text{ADC}} = 259$ counts at 50\,mm and 288 counts at 60\,mm, while the onset stays at 20 and 21\,mm. As with stretch, the loading and release strokes do not retrace the same path, the response persisting to 9 and 10\,mm on release.

\subsection{Printed Structure versus Bulk Material}

We compare the bending primitive with a plain block of the same conductive filament to verify that the sensing response comes from the printed structure rather than from the material itself. The block is printed with no internal pattern and a resting resistance of about 4\,k$\Omega$, close to the lower end of our printed primitives, so that the two samples fall within the same range of the readout circuit.

\begin{figure}[H]
  \centering
  \includegraphics[width=0.6\columnwidth]{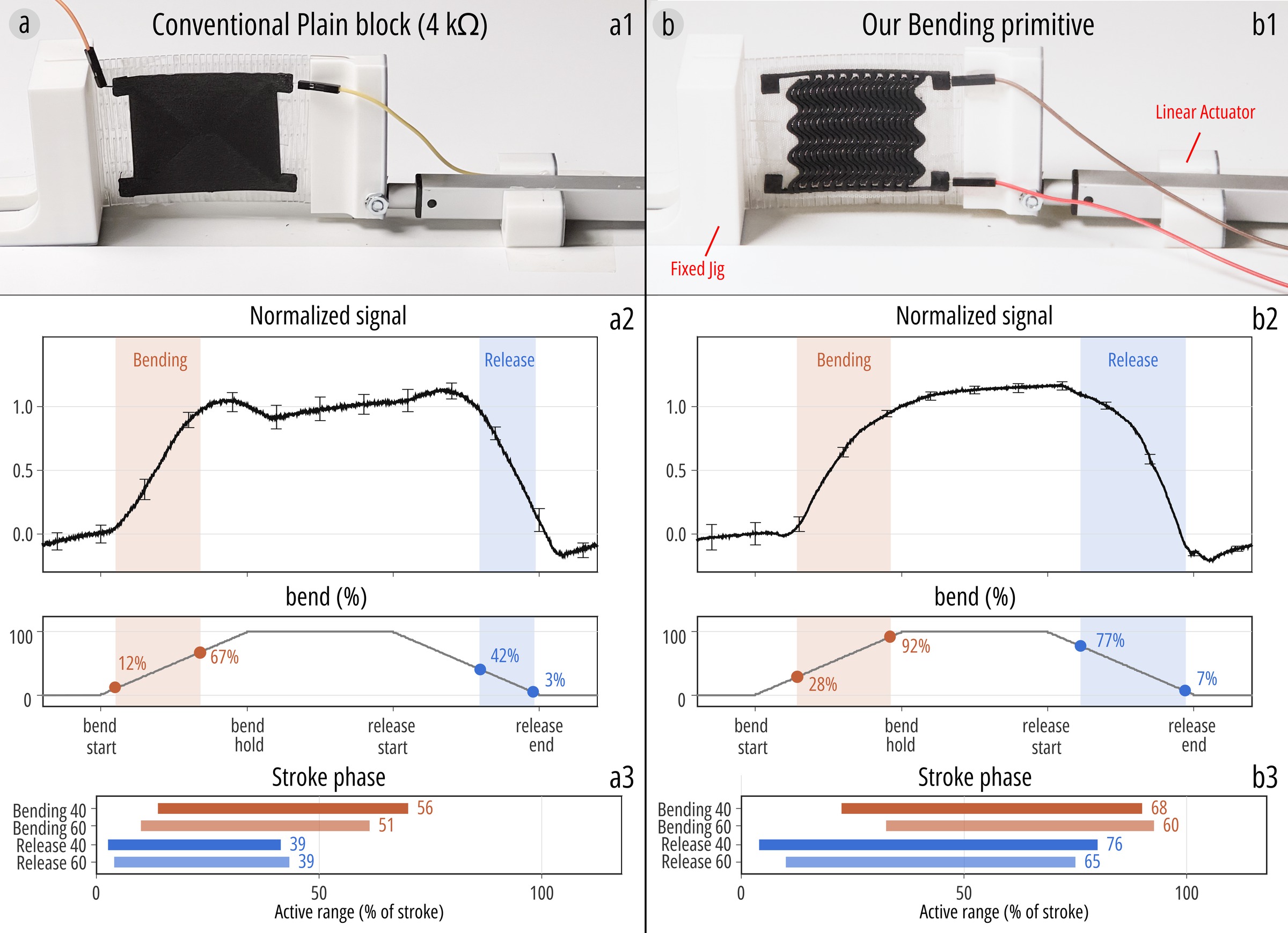}
  \caption{Difference in produced signal between conventional (a) plain block resistive sensor and (b) SenFabric bend primitive resistive sensor as function of \%bending.}

  \label{fig:bulk}
\end{figure}

\paragraph{Setup.}
Both samples are mounted in the same fixed jig and driven by the same linear actuator (Fig.~\ref{fig:bulk}, top). The actuator bends the sample to a set displacement (40\,mm or 60\,mm), holds for 3\,s, returns at the same speed, and rests flat for 3\,s. Each sample is cycled ten times; we report the last five cycles, after the initial settling of the material. The resistance is read through a voltage divider and a 12-bit ADC at 200\,Hz.

\paragraph{Active range.}
To compare where along the stroke each sample responds, we take the signal at the start of the actuator ramp as 0\% and at the end of the ramp as 100\%, and record the actuator position when the signal crosses 5\% and 95\%. We refer to the stroke between these two positions as the \emph{active range}. It is measured separately for the bending ramp and the release ramp of each cycle and expressed as a percentage of the full stroke.

\paragraph{Results.}
As shown in Fig.~\ref{fig:bulk}, the printed primitive responds over a similar portion of the stroke in both directions: 68\% while bending and 76\% while releasing at 40\,mm, and 60\% and 65\% at 60\,mm. The block responds over 56\% and 51\% of the stroke while bending, but over only 39\% while releasing at both displacements. During release the block's signal remains near its bent value while the actuator withdraws and drops only within the last 40\% of the stroke. This is expected because the resistance of the block follows the strain in the material, which relaxes slowly once the actuator no longer pushes on it; the same position therefore reads differently depending on whether the sample is being bent or released. In the printed primitive the resistance follows the contact between adjacent segments, which tracks the bend angle in both directions, so the bending and release ranges remain comparable. The primitive is therefore far less asymmetric than the block, indicating that its response is governed by the printed geometry rather than by the bulk material.

\subsection{Repeated-Use Signal Stability}

Prior work has shown that printable piezoresistive materials can exhibit history-dependent signal variation under repeated deformation~\cite{DefSense}. We therefore evaluate whether repeated actuation causes measurable degradation in SenFabric's sensing response.
We use the stretch primitive because it undergoes continuous cyclic deformation during operation, making it a representative case for evaluating repeated-use signal stability.

The sensing readout setup is identical to that used in the Signal Fingerprints evaluation. The sample is connected to an ESP32 microcontroller through two electrical contact points, and the ADC output is continuously recorded at 1~kHz during actuation. The stretch primitive is clamped at both ends on a custom 3D-printed fixture and repeatedly actuated by a linear actuator (L12-100-100-6-R, Actuonix) over a 0--25~mm stroke at approximately 9.3~mm/s, with a 1.1~s dwell at each end. Each cycle lasts approximately 7.5~s, and the sample is tested for 125 complete stretch--release cycles. All metallic fixtures in contact with the sample are electrically isolated using paper shims.

\begin{figure}[H]
  \centering
  \includegraphics[width=\columnwidth]{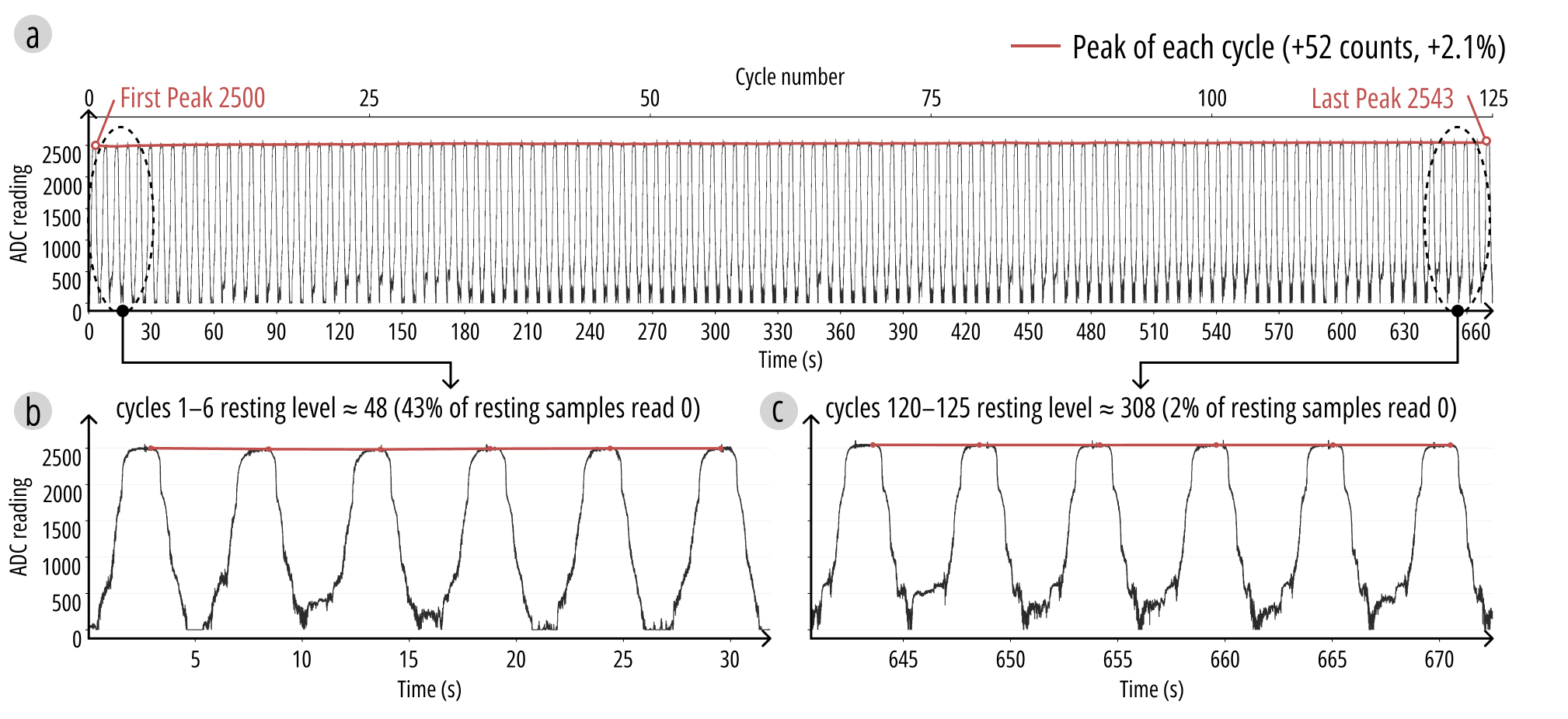}
  \caption{Signal value deviation for undeformed sensor structure across 125 cycles.}

  \label{fig:drift}
\end{figure}

As shown in Fig.~\ref{fig:drift}, repeated actuation causes a gradual increase in the resting baseline together with a modest decrease in the peak response. The baseline rises from approximately 91 counts in the initial cycles to 318 counts in the final cycles, while the peak decreases from 2506 to 2254 counts, corresponding to a change of $-10.1\%$. Despite this drift, the activated and resting states remain clearly separated throughout the test: even in the final cycles, the peak remains approximately 1936 counts above the baseline. These results indicate that SenFabric maintains a clearly distinguishable motion-induced response under repeated deformation, despite gradual baseline drift and a modest reduction in peak amplitude.

\subsection{Cross-Primitive Selectivity}\label{sec:selectivity}

To evaluate whether each sensing primitive responds selectively to its intended motion, we applied all four motion types, slide, press, bend, and stretch, to each primitive and compared the resulting responses against a primitive-specific activation threshold. The threshold was defined from the lowest response level consistently reached under the target motion. A non-target motion was considered a potential false activation if its response crossed this threshold.

\begin{figure}[H]
    \centering
    \includegraphics[width=0.6\linewidth]{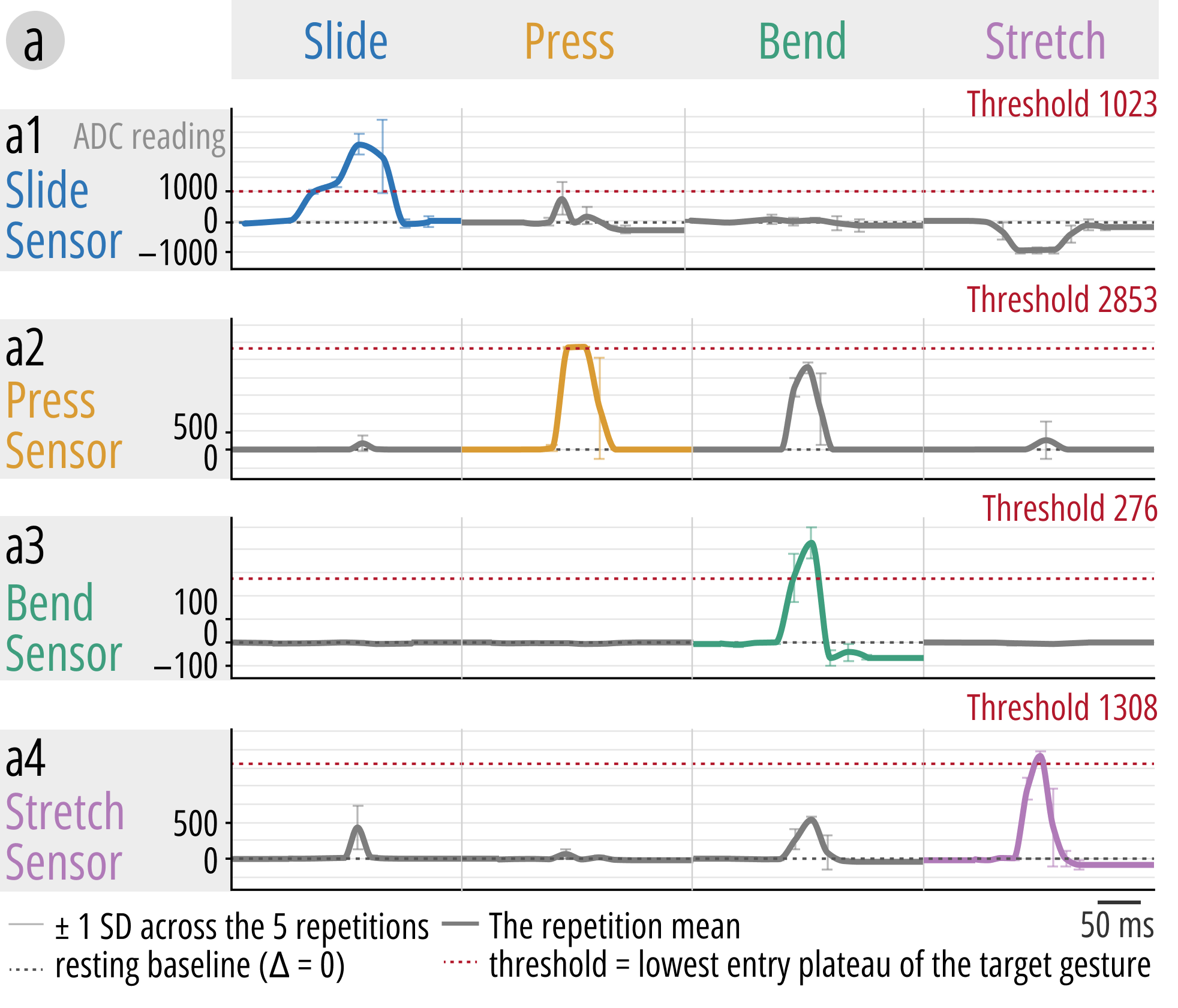}
    \caption{Cross-primitive selectivity of the four sensing primitives. Each row shows one sensor's response (mean ± 1 SD, n=5) to all four motion types, with its primitive-specific activation threshold (red dashed line). Target motions clearly cross threshold while non-target motions stay below it, except for a partial press-under-bend overlap.}
    \label{fig:selectivity}
\end{figure}

As shown in Fig.~\ref{fig:selectivity}, the slide, press, and stretch primitives show clear separation between their target and non-target motions: the target response crosses the activation threshold, while the responses produced by the other tested motions remain below it. For example, bending, pressing, or stretching the textile around the slide primitive does not reach the slide threshold and therefore is not interpreted as a sliding event. These results show that the geometry of each primitive provides substantial mechanical selectivity before signal interpretation.

The main overlap occurs for the press primitive under bending. When the press structure is placed on a highly deformable textile region, bending can deform the raised conductive feature sufficiently to cross its press activation threshold, especially for low-force press configurations. This overlap is therefore most relevant when the substrate itself undergoes substantial curvature. In applications where the press primitive is mounted on relatively rigid or low-curvature surfaces, such as robotic skin, bending-induced false activation is substantially reduced.

\section{Limitations and Future Work}

\subsection{Long-Term Durability and Washability}
The long-term durability of printed sensing structures under everyday wear has not yet been fully characterized. Repeated deformation, abrasion, laundering, and prolonged use may affect the printed geometry, adhesion, and electrical contact over time. Future work will extend fatigue testing and investigate washable fabrication strategies and protective integration methods.

\subsection{Scaling to Larger Sensor Arrays}
Our current prototypes use a small number of independently read sensing channels. Although this is sufficient for the applications demonstrated in this work, larger sensing surfaces will require more scalable wiring and readout strategies. Future work will explore multiplexed and matrix-based architectures to support denser sensor layouts with fewer electrical connections.

\subsection{Interaction Range and Placement}
The sensing primitives are designed for specific ranges of motion and interaction. Their performance can therefore depend on where they are placed on the textile and on the magnitude of deformation encountered during use. For example, regions with excessive curvature or uncontrolled fabric motion may produce responses outside the intended operating range. Future work will develop placement guidelines that match primitive geometry to the expected motion characteristics of different body locations and textile products.

\subsection{Wireless Readout and Deployment Constraints}
Several SenFabric applications already use wireless communication between the sensing hardware and host device, enabling untethered interaction in wearable settings. However, wireless deployment introduces trade-offs in communication bandwidth, latency, power consumption, and the number of sensing channels that can be streamed simultaneously. These constraints may limit applications that require high-rate sensing, very low latency, or large numbers of continuously sampled sensors. Future work will explore more efficient local processing and wireless communication strategies for these higher-demand scenarios.
\section{Conclusion}

We presented SenFabric, a design and fabrication method for directly 3D printing conductive sensing structures onto existing textiles. SenFabric uses geometry-controlled contact to reduce dependence on unstable material responses, and introduces four sensing primitives, slide, press, stretch, and bend, with tunable sensing properties and a browser-based design tool for adapting them to user-provided geometries.
Through four applications and technical evaluations, we demonstrated SenFabric across wearable sensing, human--robot interaction, garment interfaces, and animal wearables. The results show that sensing is primarily governed by printed geometry, remains selective to intended motions, and stays distinguishable under repeated deformation. SenFabric provides a flexible approach for integrating sensing directly into soft textile interfaces.

\bibliographystyle{ACM-Reference-Format}
\bibliography{sample-base}

\end{document}